\documentclass[preprint,12pt]{elsarticle} 
\usepackage{graphicx} 
\usepackage[subfigure]{graphfig}
\usepackage{epsfig}
\usepackage{epstopdf}
\usepackage{amssymb} 
\usepackage{amsmath} 

\journal{Nuclear Physics A} 

\begin{document} 

\begin{frontmatter} 

\title{From Nuclear Structure to Nucleon Structure} 
\author{Keh-Fei Liu~\footnote{liu@pa.uky.edu}} 
\address{Dept. of Physics and Astronomy \\
Center for Computational Sciences \\
 University of Kentucky, Lexington, KY 40506, USA}

\begin{abstract} 
     Similarities between nuclear structure study with many-body theory approach and nucleon structure calculations with lattice QCD are pointed out. We will give an example of how to obtain the connected sea partons from a combination of the experimental data, a global fit of parton distribution functions and a lattice calculation. We also present a complete calculation of the quark and glue decomposition of the proton
momentum and angular momentum in the quenched approximation. It is found that the 
quark orbital angular momentum constitutes about 50\% of the proton spin.

\end{abstract}
\bigskip

\end{frontmatter} 

\section{In Memoriam} 
\label{sec:ded} 

This manuscript is dedicated to the memory of Gerald E. Brown who was my Ph. D. thesis advisor,
a mentor in my professional career and a lifelong friend.  

I first met Gerry in the Fall of 1972 when I was a graduate student in Stony Brook. He just returned from NORDITA. He summoned me to his office one day and asked me if I could do some calculation for him. The problem is calculating the spectrum of two nucleons in the orbital $j$ with a delta function interaction.The next day, I went to show him my results. He had a look and said `` The gap between the $0^+$ and $2^+$ states is a factor of 2 of that between $2^+$ and $4^+$. OK, you can work for me now. '' I did not know it was a test to help him decide whether he wanted to take me on as his research assistant. 

Gerry is well known for many insightful quotations about physics. Let me relate one which is attributed to him and it may not have been recorded in a written form before. During the opening talk at one Few Body Conference, Gerry was quoted to have said ``In classical physics, you cannot solve three-body problem. With quantum mechanics, you cannot solve two-body problem and with relativistic quantum mechanics, you cannot solve one-body problem. In quantum field theory, you don't know how to solve the vacuum.'' Following Gerry's logic, we can now append his quote by ``With the advent of string theory, you no longer know where the vacuum is.''

I have learned many-body theory and Laudau's fermi-liguid theory under Gerry and my Ph. D. thesis was
on a self-consistent RPA calculation of nuclear giant resonances on Hatree-Fock ground states. In the later
years, I have followed Gerry to work on chiral soliton model of the nucleon, particularly the skyrmion.
The many intriguing properties of the nucleon both theoretically and experimentally have led me to work
on lattice quantum chromodynamics (QCD) calculation since the late eighties. 

From 1995 to 2009, we have been meeting in Caltech every January as part of a contingent of  theory guests, courtesy of Bob KcKeown and \mbox{Brad Fillipone} of the Kellogg Lab. During these visits, Gerry would explain to me his work in black holes and heavy ion collisions and I would update him on the progress in lattice QCD. Over the years, I would like to think that I have inherited part of his extraordinary enthusiasm and love for physics through osmosis and I have been influenced greatly by his way of dissecting and tackling a complex
 problem through intuition, backed by estimation. 

    It is natural to extend the study from nuclear structure to nucleon structure, especially when there is an excellent tool in lattice QCD. I am indebted to Gerry for introducing me to the fascinating world of nuclear and nucleon structures. I would like take this opportunity to  thank him for his encouragement and support over the years.

\newpage 

\section{Introduction} 
\label{sec:intro} 

      Historically, the study of nuclear structure started out from models like the liquid-drop model, 
the collective models and the shell model. The modern approaches include many-body theory, Green's function Monte Carlo and lattice effective theory calculation. Similarly, the study of nucleon structure progressed from  quark model, MIT bag model, chiral soliton model,  QCD sum rules,  instanton liquid model to the more recent lattice QCD calculation. The latter is an {\it ab initio} Euclidean path-integral calculation of QCD with controllable statistical and systematic errors. I will make a comparison between the many-body theory approach to nuclear structure and the lattice QCD approach to nucleon structure. I will draw some parallels of the two approaches and point out some differences.

\begin{figure}[tbh]
\centering
\mbox{
\hspace*{-1.1cm}
\subfigure[\label{HF}]{\includegraphics[scale=1.0]{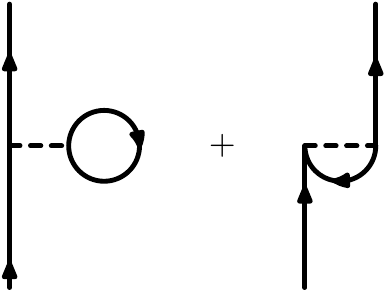}}
\hspace*{2.0cm}
\subfigure[\label{quench}]{\includegraphics[scale=1.4]{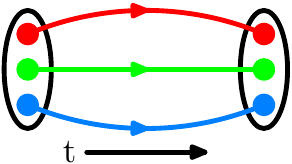}} 
}
 \caption{\small (a) Hatree-Fock diagrams and (b) quark skeleton digram of quenched QCD.}
\end{figure}

      Many-body theory is a non-relativistic quantum field theory, while QCD is a relativistic quantum field theory. As such, concepts like valence and sea degrees of freedom, collective phenomenon, and vacuum polarization are common, albeit in different contexts. In the case of nucleus, the first order of approximation is the mean-field description of the ground state of Fermi sea, such as the shell model or the Hartree-Fock approximation as depicted in Fig.~\ref{HF} and the nucleon quasi- particle and -hole states around the Fermi sea interact via an effective interaction. This is analogous to the quenched approximation of 
lattice QCD where the partition function is approximated by the gauge action only without the fermion determinant as depicted in Fig.~ref{quench}. Nucleon properties are calculated with the multi-point
correlation functions with the 3-quark interpolation field for the source and sink of the nucleon at distant time slices in the pure gauge background. 

\begin{figure}[tbh]
\centering
\mbox{
 \subfigure[\label{KB_core}]{\includegraphics[scale=1.0]{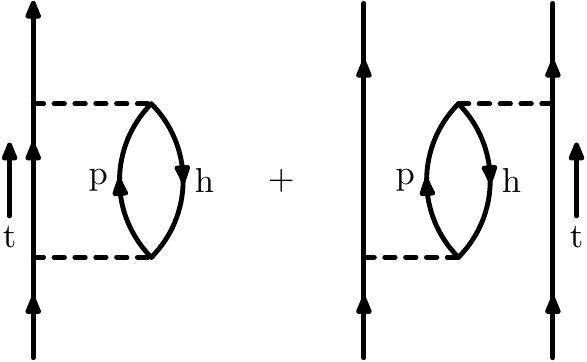}}\quad
\hspace*{2cm}
\subfigure[\label{lattice_DF}]{\includegraphics[scale=1.4]{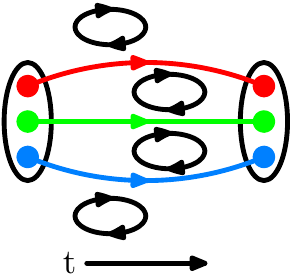}}\quad 
} 
 \caption{\small (a) Particle-phonon coupling and core excitation and (b) Lattice QCD with dynamical fermions.}
\end{figure}

     More refined approaches to nuclear structure to take into account the particle-hole excitation include single particle renormalization with particle-phonon coupling~\cite{BM75, BGG63} and Kuo-Brown interaction of the valence nucleons via core excitation of phonons~\cite{KuoBrown66}. These are illustrated
 in Fig.~\ref{KB_core}. On the nucleon structure side, the analogy would be the incorporation of the 
 dynamical fermions in the gauge background field with quark loops in the vacuum which represent the
 fermion determinant in the partition function. This is drawn schematically in Fig.~\ref{lattice_DF}.

\vspace*{0.5cm}
      
      We shall show that there are parallel developments of the same concepts in dynamics
as well as classification of degrees of freedom in many-body theory and QCD, since both are quantum field theories.  In Section~\ref{collectivity}, we shall discuss collectivity in these two theories. The Z-graph
 in nuclear structure and the corresponding connected sea partons will be compared in Section~\ref{Z-CS}.
 The core polarization will be contrasted with disconnected sea contribution in Section~\ref{core-polarization}.
 Finally, we will present the latest lattice calculation to reveal the quark and glue components of
 the proton spin in Section~\ref{spin}.

\section{Collectivity}  \label{collectivity}

\begin{figure}[tbh]
\centering
 \includegraphics[scale=0.8]{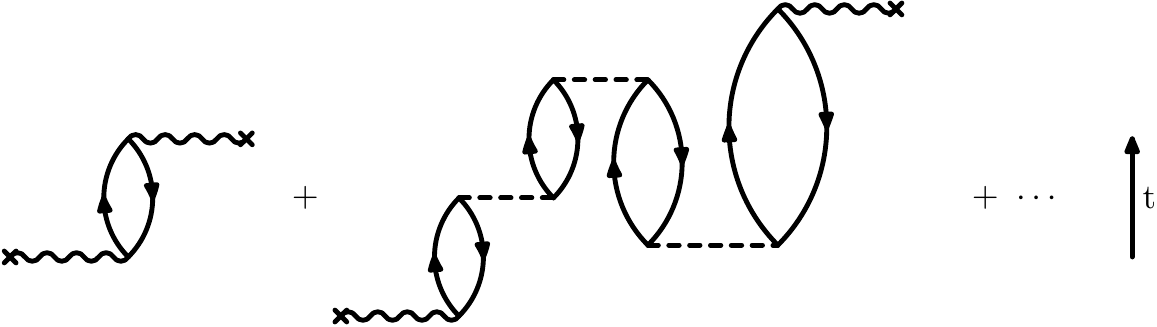}
 \caption{\small An illustration of the RPA diagrams.}
 \label{RPA}
\end{figure}

        Giant resonances in nuclei with large electric and magnetic transition rates can be 
 qualitatively understood  as a collective excitation of many particle-hole states in Gerry's schematic
 model~\cite{Brown71}. They have been successfully described in the random phase approximation (RPA) on
 Hatree-Fock ground states~\cite{LiuBrown76}. The RPA diagrams are illustrated in Fig.~\ref{RPA}.
        
\begin{figure}[tbh]
\vspace*{1cm}
\centering
\includegraphics[scale=1.4]{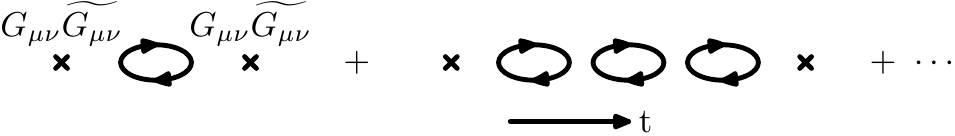}
 \caption{\small Resolution of the $U(1)$ problem in Veneziano's diagramatic approach.}
 \label{Veneziano}
 \end{figure}

In QCD, the fact that the experimental mass of $\eta'$ is much larger than all the other Nambu-Goldstone bosons 
is known as the $U(1)$ problem and is believed to be related to the $U(1)$ anomaly of the divergence of
the flavor-singlet axial current. The resolution in the context of large $N_c$ has been given
by Witten~\cite{Witten79} and Veneziano~\cite{Ven79}. In fact, the Veneziano's diagrammatic formulation as illustrated in Fig.~\ref{Veneziano} is the same as Gerry's schematic model for the degenerate particle-hole states. In the $U(1)$ case, the collective upward lift from the $u\bar{u}, d\bar{d}$ and $s\bar{s}$ `would-be' Nambu-Goldstone bosons is due to the constant coupling related to the topological susceptibility of the pure gauge theory.

We see that even though the physics contents of the giant resonance and the $U(1)$ problem are different, both are the results of collectivity as is the case of BCS superconductivity. Therefore, it is natural to employ similar formulations to tackle them.

\section{Z-graphs and Connected Sea Partons}  \label{Z-CS}

\begin{figure}[tbh]
\centering
\includegraphics[scale=1.0]{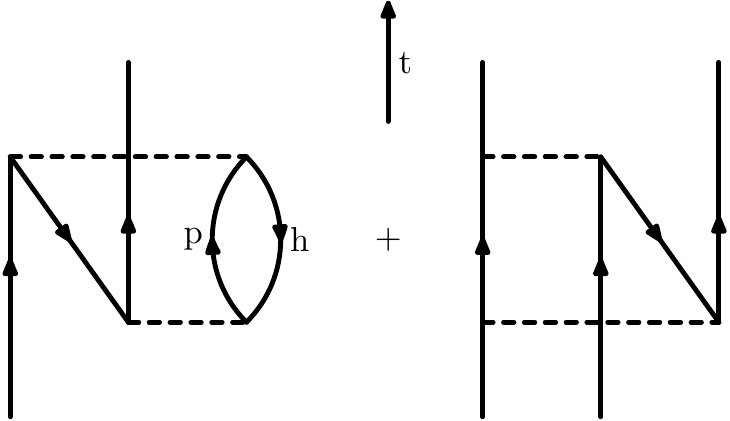}
 \caption{\small Z-graphs of the Bethe-Goldstone diagrams.}
 \label{Z}
 \end{figure}

      In many-body theory with time-ordered Bethe-Goldstone diagrams, one inevitably encounters
Z-graphs as demonstrated in Fig.~\ref{Z}. This refers to the part of the diagram where the a hole line is still connected to the valence particle lines when the interaction lines are cut. This is in contrast to the particle-hole bubble in the left diagram where the hole line is disconnected from the valence particle lines when the interaction lines are cut. 

     This distinction of the `connected hole' and `disconnected hole' diagrams had not been introduced
in the classification of the parton degrees of freedom until the surprisingly large Gottfired sum rule violation
was discovered in the deep inelastic scattering (DIS) experiment by the NMC Collaboration~\cite{nmc}. 
The  Gottfried sum rule~\cite{gottfried}, $I_G\equiv \int^1_0 dx\,[F^p_2(x)-F^n_2(x)]/x  =1/3$, was obtained
under the assumption of a symmetric $\bar u$ and $\bar d$ sea~\cite{gottfried}. The NMC measurement 
of $I = 0.235 \pm 0.026$ implies that the assumption of a symmetric $\bar u$ and $\bar d$
sea was invalid and the $x$-integrated difference of the $\bar u$ and $\bar d$ sea is 
$\int^1_0 [\bar d(x) - \bar u(x)] dx =0.148 \pm 0.039$. This striking result from the NMC was subsequently
checked using an independent experimental technique. From measurements
of the Drell-Yan cross section ratios of $(p+d)/(p+p)$, the
NA51~\cite{na51} and the Fermilab E866~\cite{e866} experiments
clearly observed the $\bar u$ and $\bar d$ difference in the
proton sea over the kinematic range of $0.015 < x < 0.35$.

\begin{figure}[hbt] \label{hadonic_tensor}
\centering
\subfigure[]
{{\includegraphics[width=0.3\hsize]{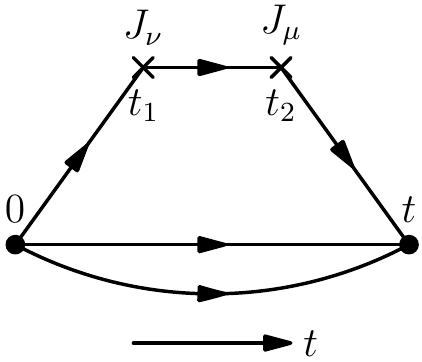}}
\label{val+CS}}
\subfigure[]
{{\includegraphics[width=0.3\hsize]{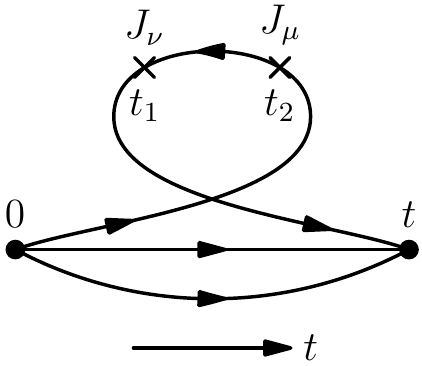}}
\label{CS}}
\subfigure[]
{{\includegraphics[width=0.3\hsize]{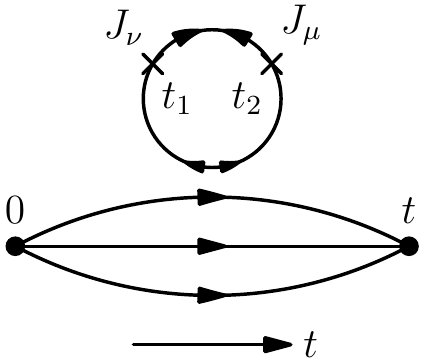}}
\label{DS}}
\caption{Three gauge invariant and topologically distinct diagrams
in the Euclidean path-integral
formulation of the nucleon hadronic tensor in the large momentum frame.
In between the currents
at $t_1$ and $t_2$, the parton degrees of freedom are
(a) the valence and CS partons $q^{v+cs}$, (b) the CS anti-partons
$\bar{q}^{cs}$, and (c) the DS partons $q^{ds}$ and
anti-partons $\bar{q}^{ds}$ with $q = u, d, s,$ and $c$. Only $u$
and $d$ are present in (a) and (b).}
\end{figure}

In order to understand the origin of this large difference between $\bar{u}$ and $\bar{d}$,
an Euclidean path-integral description of the hadronic tensor $W_{\mu\nu}$ for deep inelastic 
scattering was formulated~\cite{liu00}.  In the Bjorken limit, there are
three gauge invariant and topologically distinct diagrams, as shown in
Fig. 1. The various lines in Fig. 1 represent the quark propagators from
the source of the nucleon interpolation field at time $t=0$ to the sink time at $t$
and the currents are inserted at $t_1$ and $t_2$.

We first note that Fig.~\ref{CS}, where the quarks propagate backward in
time between $t_1$ and $t_2$, corresponds to contributions from the anti-partons which we refer as
`connected-sea' (CS) $\bar{u}^{cs}$ and $\bar{d}^{cs}$. In contrast, the time-forward propagating
quarks in Fig.~\ref{val+CS} correspond to
valence and CS partons $u^{v + cs}$ and  $d^{v + cs}$,
where the valence is defined as $q^v \equiv q^{v + cs} - \bar{q}^{cs}$
and $q^{cs}(x) \equiv \bar{q}^{cs}(x)$.
Finally, Fig.~\ref{DS} gives the the disconnected sea (DS)  $q^{ds}$ and $\bar{q}^{ds}$ for 
\mbox{$q = u,d,s,c$,} since it contains both forward and backward propagating quarks.
The nomenclature of connected and disconnected seas follows from those in
the time-ordered perturbation theory -- CS is the higher Fock-state
component in the Z-graphs where the quark lines associated with the current insertions are connected to the valence quark lines; whereas, the DS corresponds to vacuum polarization. 

We should point out a fine difference between  the CS from
Fig.~\ref{CS} and the connected hole in the Z-graphs as illustrated in Fig.~\ref{Z}. In the path-integral diagram in Fig.~\ref{CS}, the anti-quarks between the currents at $t_1$ and $t_2$ are pre-existing, i.e. they are in the nucleon wavefunction and exist before $t_1$ and after $t_2$ just as the case in Fig.~\ref{DS} so that DIS is measuring the parton density. They are not pair-produced by the hard photon from the lepton-nucleon scattering. 
The existence of the CS is easily revealed in the path-integral approach owing to the fact that it is a time-ordered formulation as are the Bethe-Goldstone diagrams. In this sense, Fig.~\ref{CS} can be considered a 
generalized Z-graph.  Similarly, Fig.~\ref{DS} can be considered a `direct diagram' in the Bethe-Goldstone sense; while Fig.~\ref{CS}, by the same token, can be considered the `exchange diagram' to reflect the fact that quarks are fermions.

 In the isospin limit where $\bar{u}^{ds}(x) = \bar{d}^{ds}(x)$, it is proved~\cite{ld95} that the DS do not contribute to the Gottfried sum rule violation. The isospin symmetry breaking is small and cannot explain the large observed violation. Rather, the majority of the violation could only come from the CS. 
 We see from Figs.~\ref{val+CS} and ~\ref{CS} that the CS and the valence are tangled together in one flavor trace. Hence,  there is no isospin symmetry between $\bar{u}^{cs}$ and $\bar{d}^{cs}$ since the state of the proton, being made up of two valence $u$ and one valence $d$, is not an isospin singlet state. Furthermore, we see that while $u$ and $d$ have both CS and DS, strange and charm have only the DS.
As far as the small-$x$ behavior is concerned, there is only reggeon exchange for the flavor non-singlet valence and CS, so the small-$x$ behavior for the valence and CS partons is
$q^{v+cs}(x),\, \bar{q}^{cs}(x) {}_{\stackrel{\longrightarrow}{x \rightarrow 0}} \propto x^{-1/2}.$
On the other hand, there is flavor-singlet pomeron exchange in addition to the reggeon exchange for the
DS partons, thus their small $x$ behaviors are $q^{ds}(x),\, \bar{q}^{ds}(x) {}_{\stackrel{\longrightarrow}{x \rightarrow 0}} \propto x^{-1}.$ Since the CS is in the same connected insertions
as the valence, it evolves like the valence; whereas, the DS evolves differently in that it has an additional 
pair-creation kernel from the gluon~\cite{liu00}. 

While the difference of $\bar{u}^{cs}(x)$ and $\bar{d}^{cs}(x)$ can be obtained from
$F_2^p(x) - F_2^n(x)$, there is not yet a well-established way to directly obtain $\bar{u}^{cs}(x) + \bar{d}^{cs}(x)$ and, for that matter, separately $\bar{u}^{cs}(x)$ and $\bar{d}^{cs}(x)$ from experiments. We have shown a way to achieve this separation with a combination of experiments, the global fit of PDF, and a lattice calculation of the momentum fraction $\langle x\rangle$ in the DI in Fig.~\ref{DI}.
The recent HERMES semi-inclusive DIS experiment of kaon production on deuteron~\cite{HERMES08} has produced the strangeness parton distribution function $s(x) + \bar{s}(x)$ at $Q^2 = 2.3 {\rm GeV}^2$. We have used this data, combined with the ratio of the lattice calculation of the strange and $u/d$ momentum fractions in the disconnected insertion, and $\bar{u}(x) + \bar{d}(x)$ from the globally fitted parton distribution function (PDF), to extract 
$\bar{u}^{cs}(x) + \bar{d}^{cs}(x)$ in the following formula
\begin{equation}  \label{udCS}
\bar{u}^{cs}(x)+\bar{d}^{cs}(x) = \bar{u}(x)+\bar{d}(x) - \frac{1}{R}(s(x) + \bar{s}(x)).
\end{equation}
at $Q^2 = 2.3\, {\rm GeV}^2$, where $\bar{u}(x)+\bar{d}(x)$ is from the CT10 PDF~\cite{CT10} which contains both the CS and DS. The strange parton distribution $s(x) + \bar{s}(x)$ is from the HERMES data and
$R$ is the ratio of the strange momentum fraction to that of the $u/d$ in the disconnected insertion in
a lattice calculation with dynamical fermions~\cite{doi08}
   \begin{equation}  \label{ratio}
     R=\frac{\langle x\rangle_{s+\bar{s}}}{\langle x\rangle_{u+\bar{u}}(DI)} = 0.857(40),
   \end{equation}
In extracting the CS in Eq.~(\ref{udCS}), we have assumed that the strange parton distribution is
proportion to that of $\bar{u}^{ds}(x)$ so that the proportional constant is $R$. In this way,
the CS $\bar{u}^{cs}(x)+\bar{d}^{cs}(x)$ is obtained in Eq.~(\ref{udCS}) by subtracting the
DS from the total $\bar{u}(x)+\bar{d}(x)$ from CT10 PDF.

\begin{figure}[htbp]
\centering
{ {\includegraphics[width=0.8\hsize]{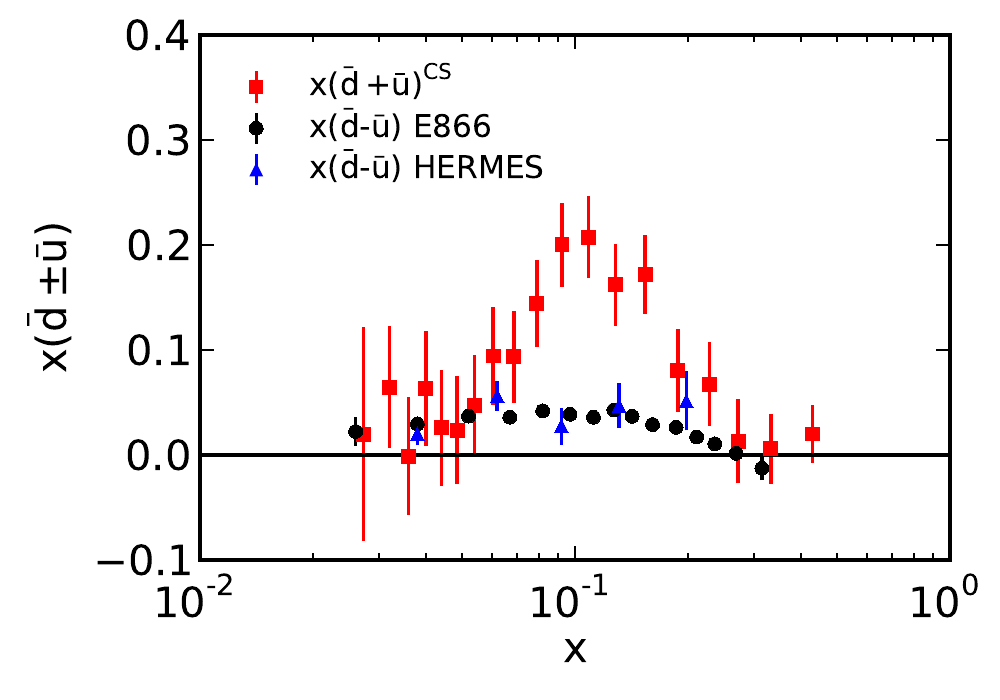}\ \ \ } }
\caption{$x(\bar{d}^{cs}(x) + \bar{u}^{cs}(x))$ obtained
from \protect{Eq.~(\ref{udCS})} is plotted together with
$x(\bar{d}(x) - \bar{u}(x))$ from E866 Drell-Yan
experiment~\protect{\cite{tma01}} and from SIDIS HERMES
experiment~\protect{\cite{ack98}}.}
\label{CSupmd}
\end{figure}

We plot the distribution function evaluated with Eq.~(\ref{udCS}), multiplied by the
momentum fraction, i.e. $x(\bar{u}(x)+\bar{d}(x) - \frac{1}{R}(s(x)
+ \bar{s}(x))$ in Fig.~\ref{CSupmd} together with $x(\bar{d}(x)-\bar{u}(x))$
from E866 Drell-Yan measurement~\cite{tma01}
at $Q^2 = 54\,\,{\rm GeV}^2$ and from semi-inclusive DIS HERMES measurement~\cite{ack98} at
$\langle Q^2\rangle = 2.3\,\,{\rm GeV}^2$.
We see that $x(\bar{u}^{cs}(x)+\bar{d}^{cs}(x))$ from Eq.~(\ref{udCS})
is peaked at medium $x \sim 0.1$, the same way as
$x(\bar{d}(x) - \bar{u}(x))$ from E866 and \mbox{HERMES}. This is consistent
with the expectation that the small-$x$
of CS, like the valence, behaves as $x^{-1/2}$ as we alluded to earlier;
so that, when CS is multiplied with $x$, it would be peaked at medium $x$,
in contrast to that of the DS, e.g. $x(s(x)+\bar{s}(s))$ from the HERMES experiment.
Furthermore, we note that $x(\bar{u}^{cs}(x)+\bar{d}^{cs}(x))$ is generally
larger than $x(\bar{d}(x) - \bar{u}(x))$ in this $x$-range as it should
and is larger by a factor $\sim$ 4 at the peak.

\begin{figure}[hbtp]
\centering
{ {\includegraphics[width=0.6\hsize]{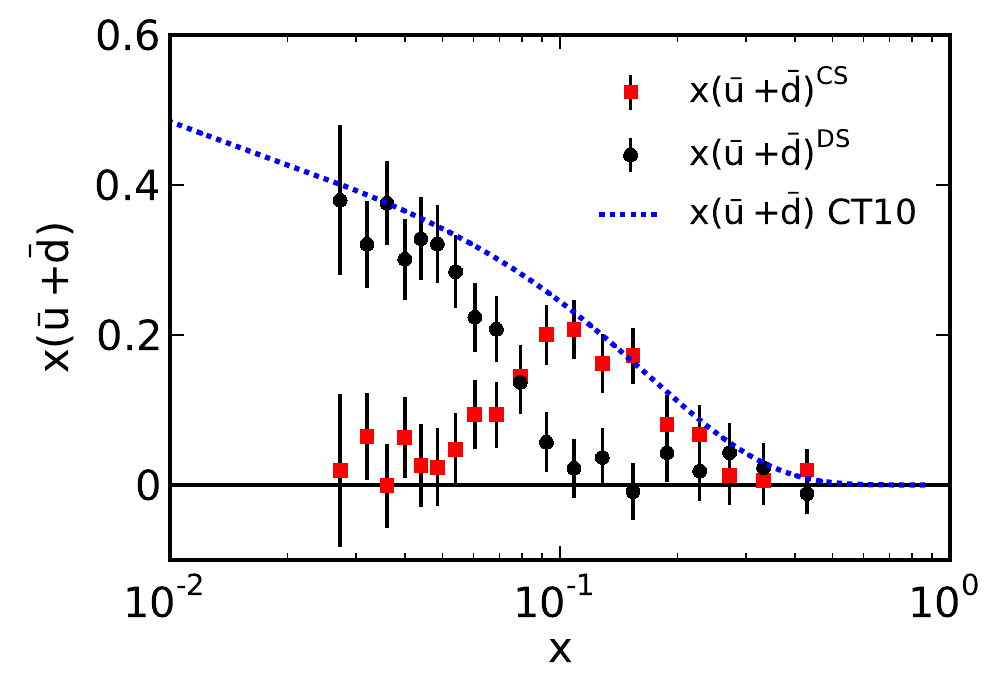}\ \ \ } }
\caption{$x(\bar{u}^{cs}(x) + \bar{d}^{cs}(x))$ obtained
from \protect{Eq.~(\ref{udCS})} is plotted together with
$x(\bar{u}(x) + \bar{d}(x))$ from CT10 and $\frac{1}{R}x(s(x)+\bar{s}(x))$
which is taken to be $x(\bar{u}^{ds}(x)+\bar{d}^{ds}(x))$.}
\label{CSu+d}
\end{figure}

We also plot $x(\bar{u}(x)+\bar{d}(x) - \frac{1}{R}(s(x) + \bar{s}(x))$,
$x(\bar{u}^{ds}(x)+\bar{d}^{ds}(x) = \frac{1}{R}x(s(x) + \bar{s}(x))$
and $x(\bar{u}(x)+\bar{d}(x))$ from CT10 in Fig.~\ref{CSu+d} to show
that the CS and DS have very different $x$-dependence. The different
shapes of CS and DS are in good agreement with the
expectation discussed earlier. This agreement lends support to the
approach we have adopted. It is interesting to note that should a very
different value of $R$ be used, the $x$-dependence of CS and DS would
no longer agree with expectation.

         Besides having different small $x$ behavior, the CS and DS evolve differently 
in the evolution equation~\cite{liu00}. The CS evolves the same way as the valence, while the
the DS evolution has an additional contribution from the gluon splitting. The global analyes
of PDFs' have not yet incorporated the separate evolutions for CS and DS.

\section{Core Polarization and Vacuum Polarization}    \label{core-polarization}

\begin{figure}[bth]
\centering
\includegraphics[scale=1.4]{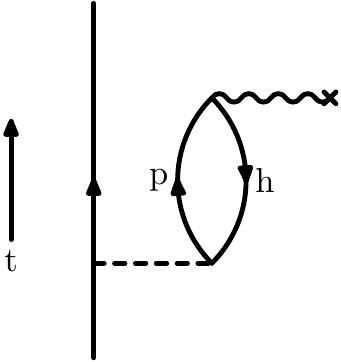}
 \caption{\small Core polarization diagram of the valance nucleon.}
 \label{core}
 \end{figure}

    In improving the magnetic moment calculation of one particle outside the closed shell, the
 core-polarization has been considered which involves the particle-hole excitation as depicted 
 in Fig.~\ref{core}. This is referred to as the Arima-Horie effect~\cite{ah54} in the nuclear physics literature,
 as shown in Fig.~\ref{core}.

\begin{figure}[hbt] \label{3-pt}
\centering
\hspace*{-1cm}
\subfigure[\label{CI}]
{\includegraphics[width=0.3\hsize]{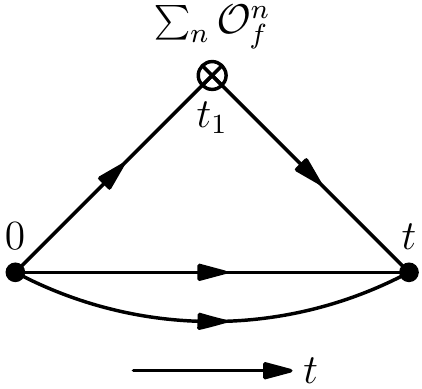}}
\hspace*{2.5cm}
\subfigure[\label{DI}]
{\includegraphics[width=0.3\hsize]{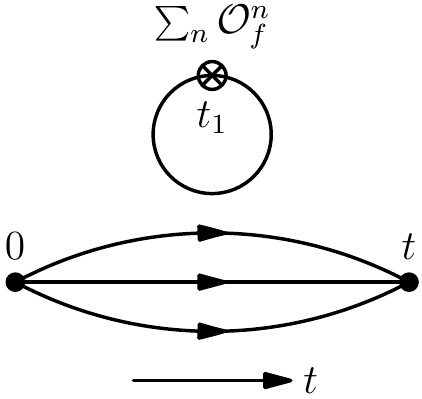}}
\caption{3-point functions for (a) connected insertions (CI) and (b) disconnected insertions (DI).}
\label{3pt}
\end{figure}

   In the case of lattice QCD, the nucleon form facts are calculated through the 3-point functions in Fig.~\ref{3pt}. The connected insertion in Fig.~\ref{CI}
contains the valence and the CS, while Fig.~\ref{DI} contains the DS. It is the latter that corresponds to the core polarization in the nuclear structure. For deep inelastic scattering, the valence and CS contributions from Fig.~\ref{val+CS} and Fig.~\ref{CS} have merged into the moment calculation of Fig.~\ref{CI} under the operator product expansion (it is the short-distance Taylor expansion in Euclidean path-integral formulation). On the
other hand, the DS parton and anti-parton contributions in Fig.~\ref{DS} has become the sum of DI moments in
Fig.~\ref{DI}.

\section{Where Does the Spin of the Proton Come from?}   \label{spin}

We have done a complete calculation of the quark and glue momenta and angular
momenta in the proton~\cite{Deka:2013zha}.\ These include the quark
contributions from both the connected and disconnected insertions.\ The quark
disconnected insertion loops are computed with $Z_4$ noise,\ and the
signal-to-noise is improved with unbiased subtractions.\ The glue operator is
comprised of gauge-field tensors constructed from the overlap operator.\ The
calculation is carried out on a $16^3 \times 24$ quenched lattice at $\beta =
6.0$ for Wilson fermions with $\kappa=0.154, 0.155$,\ and $0.1555$ which
correspond to pion masses at $650, 538$,\ and $478$~MeV,\ respectively.\ The
chirally extrapolated $u$ and $d$ quark momentum/angular momentum fraction is
found to be $0.64(5)/0.70(5)$,\ the strange momentum/angular momentum fraction
is $0.023(6)/0.022(7)$,\ and that of the glue is $0.33(6)/0.28(8)$.\ The
previous study of quark spin on the same lattice revealed that it carries a
fraction of $0.25(12)$ of proton spin.\ The orbital angular momenta of the
quarks are then obtained from subtracting the spin from their corresponding
angular momentum components.\ We find that the quark orbital angular momentum
constitutes $0.47(13)$ of the proton spin with almost all of it coming from the
disconnected insertions.

In Table~\ref{tab:chiral},\ we list the quark momentum fractions $\langle
x\rangle \equiv T_1(0)$ for the connected insertion (CI) ($u$ and $d$) and
disconnected insertion (DI) ($u/d$ and $s$) as well as that of the glue.\ We also
list the corresponding $T_2(0)$ and total angular momenta fraction $2J = T_1(0)
+ T_2(0)$ for each quark flavor and glue.\ These values are obtained at $\mu =
2$~GeV in $\overline{MS}$ scheme after perturbative renormalization and mixing
between the quark and glue operators~\cite{gl14}.

\begin{table}[htbp]
  \centering
  \renewcommand{\arraystretch}{1.4}
  \begin{tabular}{|c||cc|cccc|}
    \hline\hline
    & {\bf CI(u)} & {\bf CI(d)}  & {\bf CI(u+d)} &  {\bf DI(u/d)} & {\bf DI(s)} & {\bf Glue} \\
    \hline
    {\boldmath $\langle x \rangle$}
    & 0.416(40)  &  0.151(20) & 0.569(45) & 0.037(7) & 0.023(6) & 0.334(56) \\   
    \hline
    {\boldmath $T_2(0)$}
    & 0.287(112)  & -0.221(80) & 0.061(22) & -0.002(2) & -0.001(3) & -0.056(52) \\ 
    \hline
    {\boldmath $2J$}
    &  0.703(118)  & -0.070(82) & 0.630(51) & 0.035(7) & 0.022(7) & 0.278(76)\\  
    \hline
    {\boldmath $g_A$}
    &  0.91(11)  & -0.30(12)   & 0.62(9)  &  -0.12(1)  &  -0.12(1) & \--- \\
    \hline
    {\boldmath $2L$}
    &  -0.21(16)    &  0.23(15)   &  0.01(10)  &  0.16(1)  &  0.14(1) & \--- \\
    \hline\hline
  \end{tabular}
  \caption{Renormalized values in $\overline{MS}$ scheme at $\mu = 2$~GeV.}
  \label{tab:chiral}
\end{table}

\begin{figure}[htb]
  \centering
 {\includegraphics[width=1.0\hsize,angle=0]{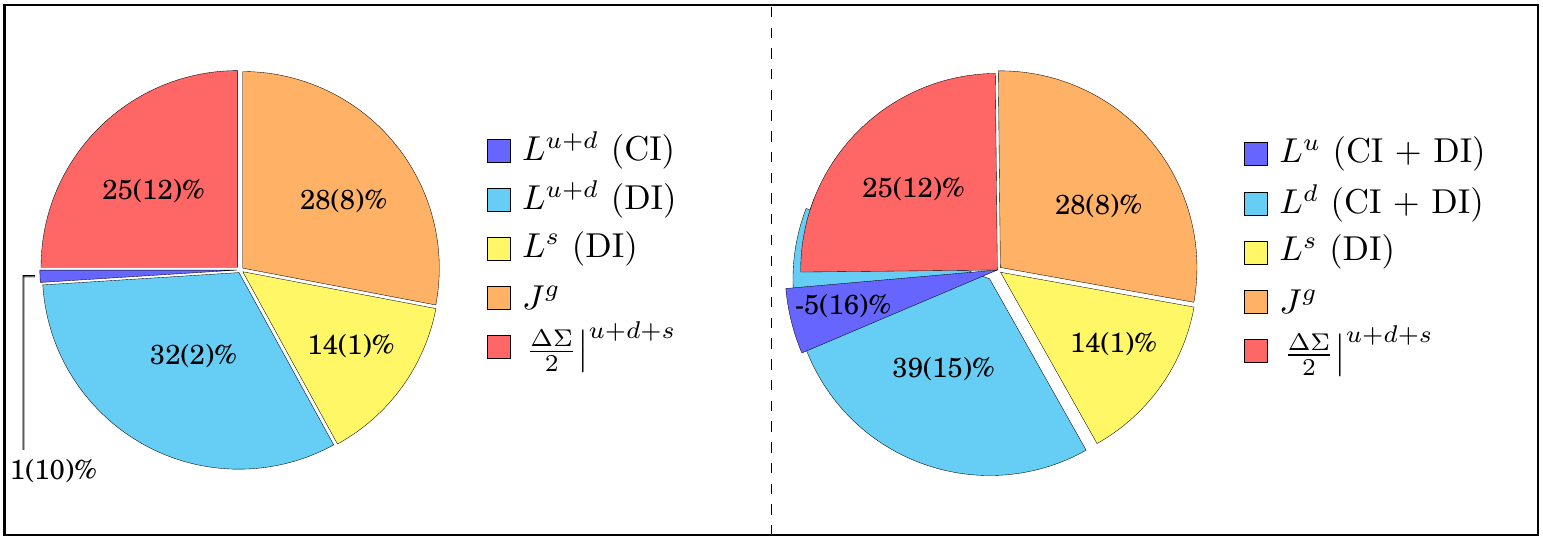}}
  \caption{Decomposition of the proton spin in terms of quark spin, glue spin,
           and quark orbital angular momentum for the $u,d$ and $s$ quarks.}
  \label{pie}
\end{figure}

We illustrate the composition of proton spin in a pie chart in
Fig.~\ref{pie}. The total quark spin constitutes 25(12)\% and the glue also
gives 28(8)\%. The rest (47(13)\%) comes from the quark orbital angular
momentum. Since the $u$ and $d$ quark orbital angular momenta in the connected
insertion (i.e.\ valence) almost cancel, almost all of the quark orbital
angular momentum comes from the disconnected insertion (i.e.\ vacuum
polarization). We have used the momentum and angular momentum sum rules to
obtain the renormalization constants for the lattice quark and glue
energy-momentum tensor operators and used perturbative renormalization and
mixing between the quark and glue operators to quote our final numbers in the
$\overline{MS}$ scheme at $\mu = 2$~GeV.  This is the first time that a complete
calculation on the decomposition of the proton momentum and spin is carried out.

\subsection{Quark spin from anomalous Ward identity}

Now that we have gone through the complete calculation in the quenched
approximation, the next step in to carry out the same calculations with valence
overlap fermion on $2+1$-flavor dynamical domain-wall fermion gauge
configurations on several lattices with different lattice spacings and sea quark
masses so that we can obtain definitive results of the proton spin
components at the physical pion mass and  the continuum limit with realistic
dynamical quarks in the vacuum in order to compare with experiments and make
predictions. 

One of the quantities to calculate is the quark spin and the most challenging
part is the disconnected insertion. We first calculated the quark loop for the
axial-vector current and found that, contrary to the pseudoscalar and scalar
cases, it is not dominated by the low modes. As a result, one needs a large
number of noises to control the statistical error of the high modes. Instead of
increasing the number of noises, we take another approach.  We shall calculate the
axial-vector matrix element from the anomalous Ward identity
\begin{equation}\label{awi2}
\partial^{\mu} A_{\mu}^0 = 2 \sum_{f=1}^{N_f} m_f \overline{q}_f
 \gamma_5 q_f + N_f  2 q ,
\end{equation}
where $q$ is the local topological charge operator. For the overlap fermion, 
the topological charge is given by the overlap Dirac operator, i.e.  \\
\mbox{$q(x) = Tr \gamma_5( 1 - \frac{1}{2}
D_{ov} (x,x))$.}  We see from the r.h.s of Eq.~(\ref{awi2}) that the
renormalization of the pseudoscalar density is canceled by the renormalization
of the quark mass for overlap fermion and there is no renormalization for the
topological term due the index theorem as obeyed by the $q$ defined by the
overlap operator. Therefore, calculating the r.h.s. of Eq.~(\ref{awi2}) gives
the non-perturbatively renormalized axial-vector matrix element which is the
quark spin.  We show our preliminary results on the topological charge
contribution whose signal is coded in the slope of the figure.

\begin{figure}[htb]
  \centering
 {\includegraphics[width=0.6\hsize,angle=0]{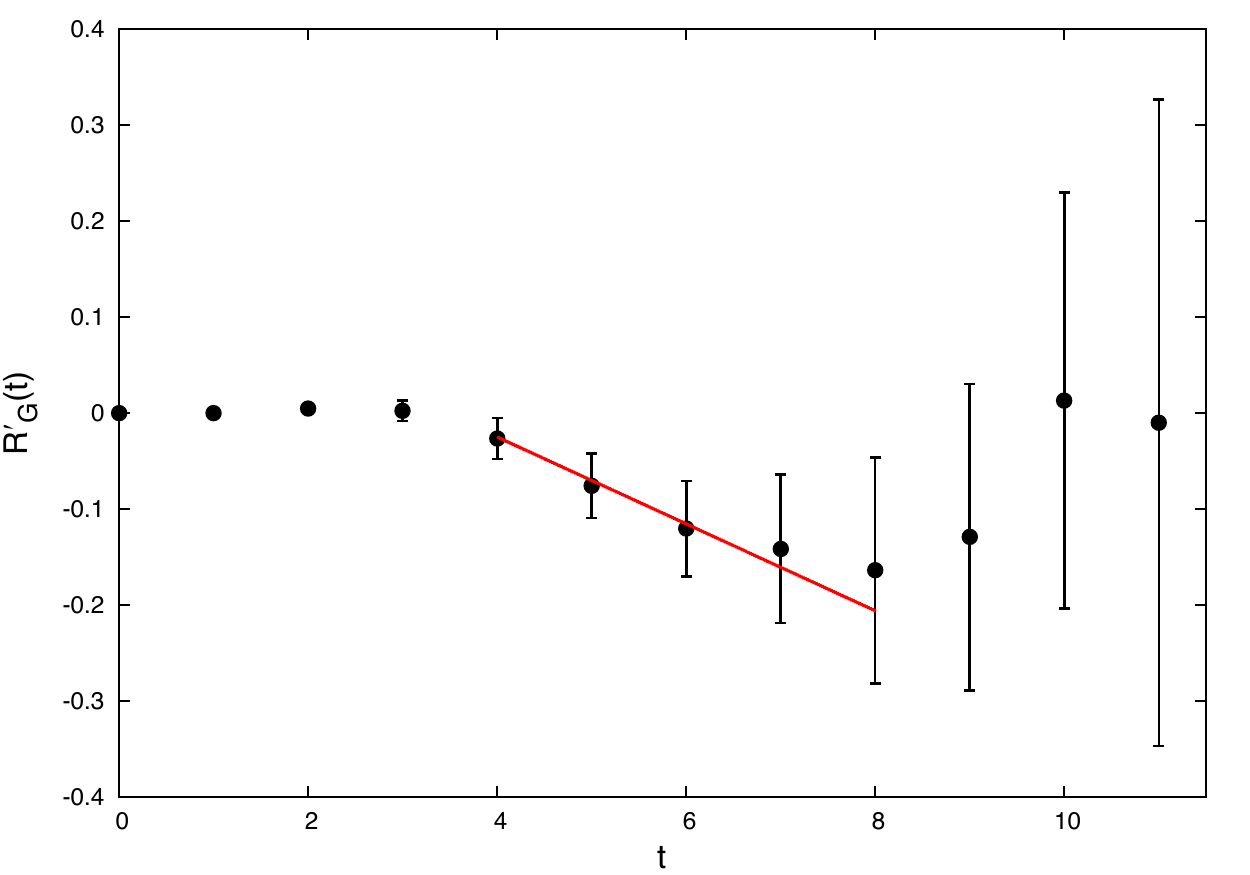}}
  \caption{The three-point to two-point function ratio in the sum method  as a
                function of time where the anomaly contribution is the slope of the ratio 
                after the nucleon appears in the two-point correlator. This is at $q^2 = -0.186\, {\rm GeV}^2$.}
  \label{ratio}
\end{figure}

The slope is $-0.045(16)$. This is done on the $24^3 \times 64$ lattice with
100 configurations and both the valence and sea quarks are at pion mass of 330
MeV. This is the contribution of the anomaly to the quark spin for each flavor
and at $q^2 = - 0.186 {\rm GeV}^2$. When the $q^2$ extrapolation to $q^2 = 0$
and chiral extrapolation to the physical pion are carried out, we expect the
number to be larger.  The pseudoscalar density contribution turns out to the
very small. We are analyzing them with more configurations and with more
nucleon sources. A statistical error of 15\% is targeted.  Production is
ongoing for the two $32^3 \times 64$ lattices with lattice spacings of $0.085$
and $0.14$ fm. When the continuum and physical pion mass limits are taken, we
should have a better picture on the quark spin in the proton. 

\section{Summary}

      We see that nuclear structure in many-body theory and lattice QCD calculation
of nucleon structure share many conceptual similarities in the classification of fermion degrees of
freedom and collective phenomena. However, there is a significant difference in that the nucleon structure involves explicit glue contributions which are absent in the study of nuclear structure. Furthermore, there are triangle anomaly and trace anomaly in QCD which have significant consequences in quark spin and nucleon mass. These are novel and challenging features in the study of nucleon structure both theoretically and experimentally.

\section*{Acknowledgments} I thank my colleagues in the $\chi QCD$ Collaboration for the lattice
results. The research is supported partially by the U.S. Department of Energy grant DE-FG05-84ER40154 and the Center for Computational Sciences of University of Kentucky.


\begin{thebibliography}{99} 


\bibitem{BM75}
A. Bohr and B.R. Mottelson, {\it Nuclear Structure}, Vol. II, pp. 416  (W.A. Benjamin, Inc. 1975).

\bibitem{BGG63}
G.E. Brown, J.H. Gunn, P. Gould, Nucl. Phys., {\bf 46}, 598 (1963).

\bibitem{KuoBrown66}
T.T.S. Kuo and G.E. Brown, Nucl. Phys. {\bf 85}, 40 (1966).

\bibitem{Brown71}
G.E. Brown, Chap. V. 3 in {\it Unified Theory of Nuclear Models and Forces} (North-Holland Pub. Co. 
Amsterdam 1971). 

\bibitem{LiuBrown76}
K.F. Liu and G.E. Brown, Nucl. Phys. {\bf A265}, 385 (1976).

\bibitem{Witten79}
E. Witten, Nucl. Phys. {\bf B156}, 269 (1979).

\bibitem{Ven79}
 G. Veneziano, Nucl. Phys. {\bf B159}, 213 (1979); Phys. Lett. {\bf 95 B}, 90 (1980).

\bibitem{nmc} A. Amaudruz {\it et al.} (NMC Collaboration),
Phys. Rev. Lett. {\bf 66}, 2712 (1991); M. Arneodo {\it et al.},
Phys. Rev. D {\bf 50}, R1 (1994).

\bibitem{gottfried} K. Gottfried, Phys. Rev. Lett.
{\bf 18} 1174 (1967).

\bibitem{na51} A. Baldit {\it et al.} (NA51 Collaboration),
Phys. Lett. B {\bf 332}, 244 (1994).

\bibitem{e866} E. A. Hawker {\it et al.} (E866/NuSea
Collaboration), Phys. Rev. Lett. {\bf 80}, 3715 (1998);
J. C. Peng {\it et al.}, Phys. Rev. D {\bf 58}, 092004 (1998);
R.S. Towell {\it et al.}, Phys. Rev. D {\bf 64}, 052002 (2001).

\bibitem{liu00} K.F. Liu, Phys. Rev. {\bf D62}, 074501 (2000).

\bibitem{ld95}
K.F. Liu and S.J. Dong, Phys. Rev. Lett. {\bf 72}, 1790 (1994).

\bibitem{HERMES08}
A. Airapetian et al. (HERMES), Phys. Lett. {\bf B666}, 446 (2008).

\bibitem{CT10}
H.L. Lai, M. Guzzi, J. Huston, Z. Li, P.M. Nadolsky, J. Pumplin, C.-P. Yuan,
Phys. Rev. {\bf D82}, 074024 (2010), [arXiv:1007.2241].


\bibitem{doi08}
T. Doi {\it et al.} ($\chi$ QCD Collaboration), PoS {\bf LATTICE2008}, 163 (2008), [arXiv:0810.2482].

\bibitem{tma01}
R.S. Towell {\it et al.} (FNAL E866/NuSea Collaboration), Phys. Rev. {\bf D64}, 052002 (2001).

\bibitem{ack98}
A. Ackerstaff {\it et al.} (HERMES), Phy. Rev. Lett. {\bf 81}, 5519 (1998).

\bibitem{ah54}
A. Arima and H. Horie, Prog. Theor. Phys. {\bf 12}, 623 (1954).

\bibitem{Deka:2013zha} 
  M.~Deka, T.~Doi, Y.B.~Yang, B.~Chakraborty, S.J.~Dong, T.~Draper, M.~Glatzmaier 
  and M.~Gong,  H.W.~Lin, K.F.~Liu, D.~Mankame, N.~Mathur, and T.~Streue,
  arXiv:1312.4816 [hep-lat].
  
\bibitem{gl14}
Michael Glatzmaier and K.F. Liu, arXiv:1403.7211.
  
  

\end{thebibliography}
\end{document}